\documentclass[aps,prl,twocolumn,amsmath,amssymb,citeautoscript,superscriptaddress]{revtex4-1}
\usepackage{graphicx}
\usepackage{dcolumn}
\usepackage{bm}
\usepackage{accents}

\newcommand{\aap}{Astron.\ Astrophys.}

\newcommand{\apjl}{Astrophys.\ J.\ Lett.}

\newcommand{\araa}{Ann.\ Rev.\ Astron.\ Astrophys.}

\newcommand{\cntext}[1]{~}

\begin{document}

\title{Gravitational Test Beyond the First Post-Newtonian Order with the Shadow of the M87 Black Hole}

\author{Dimitrios Psaltis}
\affiliation{Steward Observatory and Department of Astronomy, University of Arizona, 933 N. Cherry Ave., Tucson, AZ 85721, USA}

\author{Lia Medeiros}
\affiliation{School of Natural Sciences, Institute for Advanced Study, 1 Einstein Drive, Princeton, NJ 08540, USA}

\author{Pierre Christian}
\affiliation{Steward Observatory and Department of Astronomy, University of Arizona, 933 N. Cherry Ave., Tucson, AZ 85721, USA}

\author{Feryal Özel}
\affiliation{Steward Observatory and Department of Astronomy, University of Arizona, 933 N. Cherry Ave., Tucson, AZ 85721, USA}

\author{Kazunori Akiyama}
\affiliation{National Radio Astronomy Observatory, 520 Edgemont Rd, Charlottesville, VA 22903, USA}
\affiliation{Massachusetts Institute of Technology Haystack Observatory, 99 Millstone Road, Westford, MA 01886, USA}
\affiliation{National Astronomical Observatory of Japan, 2-21-1 Osawa, Mitaka, Tokyo 181-8588, Japan}
\affiliation{Black Hole Initiative at Harvard University, 20 Garden Street, Cambridge, MA 02138, USA}

\author{Antxon Alberdi}
\affiliation{Instituto de Astrof\'{\i}sica de Andaluc\'{\i}a-CSIC, Glorieta de la Astronom\'{\i}a s/n, E-18008 Granada, Spain}

\author{Walter Alef}
\affiliation{Max-Planck-Institut f\"ur Radioastronomie, Auf dem H\"ugel 69, D-53121 Bonn, Germany}

\author{Keiichi Asada}
\affiliation{Institute of Astronomy and Astrophysics, Academia Sinica, 11F of Astronomy-Mathematics Building, AS/NTU No. 1, Sec. 4, Roosevelt Rd, Taipei 10617, Taiwan, R.O.C.}

\author{Rebecca Azulay}
\affiliation{Departament d'Astronomia i Astrof\'{\i}sica, Universitat de Val\`encia, C. Dr. Moliner 50, E-46100 Burjassot, Val\`encia, Spain}
\affiliation{Observatori Astronòmic, Universitat de Val\`encia, C. Catedr\'atico Jos\'e Beltr\'an 2, E-46980 Paterna, Val\`encia, Spain}
\affiliation{Max-Planck-Institut f\"ur Radioastronomie, Auf dem H\"ugel 69, D-53121 Bonn, Germany}

\author{David Ball}
\affiliation{Steward Observatory and Department of Astronomy, University of Arizona, 933 N. Cherry Ave., Tucson, AZ 85721, USA}

% ...

\collaboration{the EHT Collaboration}
\noaffiliation


\begin{abstract}
The 2017 Event Horizon Telescope (EHT) observations of the central source in M87 have led to the first measurement of the size of a black-hole shadow. This observation offers a new and clean gravitational test of the black-hole metric in the strong-field regime. We show analytically that spacetimes that deviate from the Kerr metric but satisfy weak-field tests can lead to large deviations in the predicted black-hole shadows that are inconsistent with even the current EHT measurements. We use numerical calculations of regular, parametric, non-Kerr metrics to identify the common characteristic among these different parametrizations that control the predicted shadow size. We show that the shadow-size measurements place significant constraints on deviation parameters that control the second post-Newtonian and higher orders of each metric and are, therefore, inaccessible to weak-field tests. The new constraints are complementary to those imposed by observations of gravitational waves from stellar-mass sources.
\end{abstract}


\maketitle

Tests of general relativity have traditionally involved solar-system bodies~\cite{Will2014} and neutron stars in binaries~\cite{Wex2014}, for which precise measurements can be interpreted with minimal astrophysical complications. In recent years, observations at cosmological scales~\cite{Ferreira2019} and the detection of gravitational waves~\cite{Abbott2016,*Abbott2019,*Ilsi2019} have also resulted in an array of new gravitational tests.

The horizon-scale images of the black hole in the center of the M87 galaxy obtained by the EHT~\cite{PaperI,*PaperII,*PaperIII,*PaperIV,*PaperV,*PaperVI} offer the most recent addition to the set of observations that probe the strong-field regime of gravity. As an interferometer, the EHT measures the Fourier components of the brightness distribution of the source on the sky at a small number of distinct Fourier frequencies. The features of the underlying image are then reconstructed either using agnostic imaging algorithms or by directly fitting model images to the interferometric data. The central brightness depression seen in the M87 image has been interpreted as the shadow cast by this supermassive black hole on the emission from the surrounding plasma. The observability of the shadow of the black hole in M87 and the one the center of the Milky Way, Sgr~A*, had been predicted earlier based on the properties of the radiatively inefficient accretion flows around these objects and their large mass-to-distance ratios~\cite{Falcke2000,*Ozel2000,*Ozel2001}. 

The outline of a black-hole shadow is the locus of the photon trajectories on the screen of a distant observer  that, when traced backwards, become tangent to the surfaces of spherical photon orbits hovering just above the black-hole horizons~\cite{Bardeen1973,*Chandra1983,*Tao2003,*Takahashi2004}. The Boyer-Lindquist radii of these spherical photon orbits lie in the range $(1-4) M$, depending on the black-hole spin and the orientation of the angular momentum of the orbit~\cite{Bardeen1972} (here $M$ is the mass of the black hole and we have set $G=c=1$, where $G$ is the gravitational constant, and $c$ is the speed of light). It is the fact that the outlines of black-hole shadows encode in them the strong-field properties of the spacetimes that led to the early suggestion that they can be used in performing strong-field gravitational tests~\cite{Bambi2009, Johannsen2010, Psaltis2019,*Cunha2018,*Baker2015}.

Even though the radii of the photon orbits have a strong dependence on spin, a fortuitous cancellation of the effects of frame dragging and of the quadrupole structure in the Kerr metric causes the outline of the shadow, as observed at infinity, to have a size and a shape that depends very weakly on the spin of the black hole or the orientation of the observer~\cite{Johannsen2010}. This cancellation occurs because, due to the no-hair theorem, the magnitude of the quadrupole moment of the Kerr metric is not an independent quantity but is instead always equal to the square of the black-hole spin. For all possible values of spin and inclination, the size of the shadow is $\simeq 5 M\pm 4$\% and its shape is nearly circular to within $\sim 7$\%. For a black hole of known mass-to-distance ratio, the constancy of the shadow size allows for a null-hypothesis test of the Kerr metric~\cite{Psaltis2015}. At the same time, the nearly circular shape of the shadow offers the possibility of testing the gravitational no-hair theorem~\cite{Johannsen2010}.

The first inference of the size of the black-hole shadow in M87 used as a proxy the measurement of the size of the bright ring of emission that surrounds the shadow and calibrated the difference in size via large suites of GRMHD simulations~\cite{PaperV}. When this ring of emission is narrow, as is the case for the 2017 EHT image of M87, potential biases in the measurement are small. The inferred size of the M87 black-hole shadow was found to be consistent (to within $\sim 17$\% at the 68-percentile level) with the predicted size based on the Kerr metric and the mass-to-distance ratio of the black hole derived using stellar dynamics~\cite{PaperVI,Gebhardt2011} (see, however, ~\cite{Walsh2013,Kormendy2013} and \footnote{A second technique of measuring the mass of M87 based on gas dynamics results in a mass that is smaller by a factor of 2. However, the accuracy of this technique has been questioned as it often leads to underestimated black-hole masses~[15]. Moreover, choosing this mass instead would lead us to the conclusion that the metric of the M87 black hole deviates substantially from Kerr. We consider this to be a rather unlikely possibility and, therefore, give a negligible prior to the mass measurement from gas dynamics}). The agreement between the measured and predicted shadow size does constitute a null-hypothesis test of the GR predictions: the data give us no reason to question the validity of the assumptions that went into this measurement, the Kerr metric being one of them. However, using this measurement to place quantitative constraints on any potential deviations from the Kerr metric is less straightforward for two reasons.

First, the Kerr metric is the unique black-hole solution to a large number of modified gravity theories that are Lorentz symmetric and have field equations with constant coupling coefficients between the various gravitating fields~\cite{Psaltis2008b,Sotiriou2012}. Only a limited number of black-hole solutions are known for theories with dynamical couplings~\cite{Berti2015} (e.g., dynamical Chern-Simons gravity and Einstein-dilaton-Gauss-Bonnet gravity~\cite{Yunes2011,*Yagi2012,*Ayzenberg2014,*McNees2016,*Silva2018,*Doneva2018}) or for Lorentz-violating theories~\cite{Barausse2011,*Barausse2013,*Barausse2016,*Ramos2019}.  Despite substantial progress in recent years, this line of work leads to limited theoretical guidance on the form and magnitude of potential deviations from the Kerr metric.

Second, if we instead use an empirical parametric framework to extend the Kerr metric, we would find that most naive parametric extensions lead to pathologies, such as non-Lorentzian signatures, singularities, and closed timelike loops, which render it impossible to calculate photon trajectories in the strong-field regime~(see, e.g.,~\cite{Johannsen2013}). In recent years, this problem has been addressed with the development of a number of parametric extensions of the Kerr metric that are free of pathologies~\cite{Johannsen2011,Vigeland2011,Johannsen2013b,Cardoso2014,Rezzolla2014,Cardoso2015,Konoplya2016}. Resolving the pathologies, however, comes at the cost of very large complexity. In principle, we can use the EHT measurement with any of these parametric extensions to place constraints on the specific parameters of the metric we used~\cite{Johannsen2016}. However, understanding the physical meaning of such constraints and comparing them with the constraints imposed when other parametric extensions are used are not readily feasible. In addition, the complexity of the various parametric extensions to the Kerr metric hinders the comparison of these gravitational tests with the results of other, e.g., weak-field and cosmological ones and, therefore, the effort to place complementary tests on the underlying gravity theory.

In this {\em Letter}, we  use analytic arguments as well as numerical calculations of shadows to set new constraints on gravity using the 2017 EHT measurements, elucidate their physical meaning, and compare them with earlier weak-field tests. We find that the EHT measurements place constraints primarily on the $tt-$element of the black-hole spacetime (when the latter is expressed in areal coordinates and in covariant form). This is analogous to the fact that solar-system tests that involve gravitational lensing or Shapiro delay measurements constrain primarily one of the metric elements of the PPN framework~\cite{Will2014}. However, we show that the constraints imposed by the EHT measurements are of (at least) the second post-Newtonian order and are, therefore, beyond the reach of weak-field experiments.

The size of the black-hole shadow both in the Kerr metric and in other parametric extensions depends very weakly on the black-hole spin~\cite{Johannsen2010,Johannsen2013c,Medeiros2020}. For this reason, we start by exploring analytically the shadow size for a general static, spherically symmetric metric of the form
\begin{equation}
ds^2=g_{tt} dt^2 + g_{rr} dr^2 + r^2d\Omega\;.
\label{eq:metric_gen}
\end{equation}
Note that the choice of coordinates we use here is different from the isotropic coordinates of the PPN framework. We made this choice because, as we will show below, the radius of the photon orbit and the size of the shadow depend only on this element of the metric in these coordinates (unlike, e.g., Eq.~[101] of Ref.~\cite{Sullivan2020}, which is written in isotropic coordinates).

Without loss of generality, we consider photon trajectories in the equatorial plane, i.e., set $\theta=\pi/2$.
Following Ref.~\cite{Psaltis2008}, we use two of the Killing vectors of the spacetime to write the components of the momentum of a photon traveling in this spacetime as
\begin{equation}
(k^t,k^r, k^\theta, k^\phi)=\left( \frac{E}{g_{tt}},
\left[-\frac{E^2}{g_{tt}\;g_{rr}}-\frac{l^2}{g_{rr}r^2}\right]^{1/2},
0, \frac{l}{r^2}\right)\;,
\end{equation}
where $E$ and $l$ are the conserved energy and angular momentum of the photon and we have
used the null condition $\underaccent{\tilde}{k}\cdot \underaccent{\tilde}{k}=0$ to calculate the radial component of the momentum.

\begin{table}[t]
\caption{PPN expansions of various parametric extensions to the Kerr metric}
\label{tab:table1}
\begin{ruledtabular}
\begin{tabular}{lcc}
\textrm{Metric}&
\textrm{$\bar{\beta}-\bar{\gamma}$} (1PN) &
\textrm{$\zeta$} (2PN)\\
\colrule
Kerr & 0 & 0 \\
JP & 0 & $\alpha_{13}$ \\
MGBK & $-\gamma_{1,2}/2-\gamma_{4,2}\rightarrow 0$ & $-\gamma_{1,2}-4\gamma_{4,2}\rightarrow \gamma_{1,2}$ \\
\end{tabular}
\end{ruledtabular}
\end{table}

The location of the circular photon orbit is the solution of the two conditions $k^r=0$ and $dk^r/dr=0$. Combining them, we write the radius $r_{\rm ph}$ of the photon orbit as the solution to the implicit equation
\begin{equation}
r_{\rm ph}=\sqrt{-g_{tt}} \left(\left.\frac{d\sqrt{-g_{tt}}}{dr}\right\vert_{r_{\rm ph}}\right)^{-1}\;.
\end{equation}
The radius $r_{\rm sh}$ of the black-hole shadow as observed at infinity is the gravitationally lensed image of the circular photon orbit. This effect was calculated in Ref.~\cite{Psaltis2008} (Eq.~[20]) and, when applied to the size of the photon orbit, leads to
\begin{equation}
r_{\rm sh}=\frac{r_{\rm ph}}{\sqrt{-g_{tt}(r_{\rm ph})}}\;.
\label{eq:rsh}
\end{equation}
As advertised earlier, both the radius of the photon orbit and the size of the black-hole shadow depend only on the $tt$ element of the metric~(\ref{eq:metric_gen}) written in areal coordinates and in covariant form.

\begin{figure*}[t]
\includegraphics[width=0.43\textwidth]{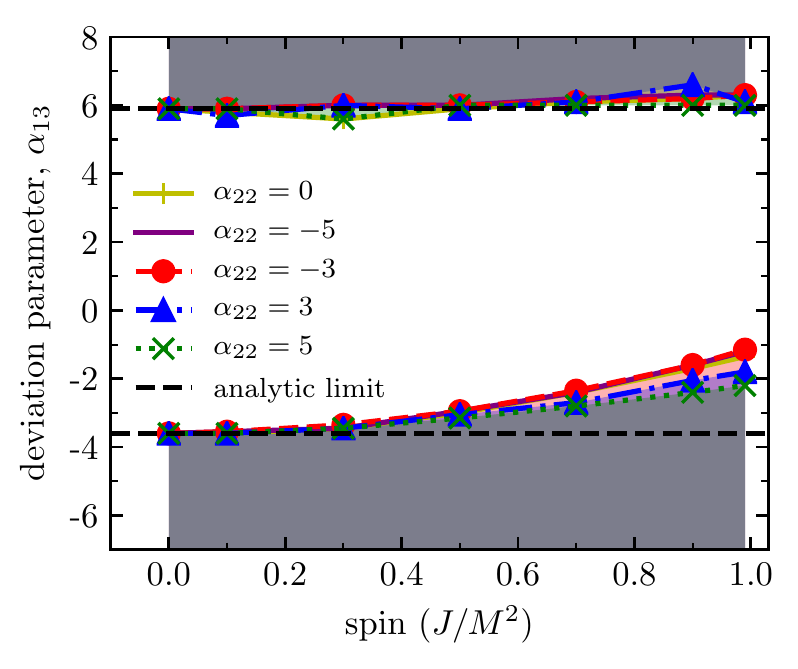}
\includegraphics[width=0.43\textwidth]{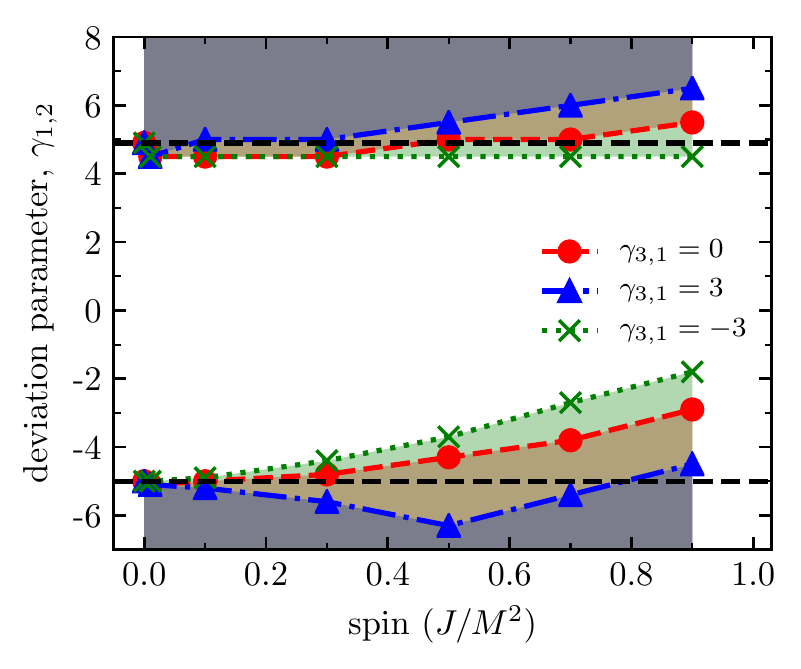}
\caption{\label{fig:metrics}  Bound on the deviation parameters {\em (Left)\/} $\alpha_{13}$ of the JP metric and {\em (Right)\/} $\gamma_{1,2}$ for the MGBK metric, as a function of spin ($J/M^2$) and for different values of the other metric parameters, placed by the 2017 EHT observations of M87. The shaded areas show the excluded regions of the parameter space. The dashed line shows the analytic result obtained for zero spin. The EHT measurements place constraints predominantly on $\alpha_{13}$ (for JP) and $\gamma_{1,2}$ (for MGBK), which control the 2PN expansion of the corresponding metrics (see Table~\ref{tab:table1}).}
\end{figure*}

In order to connect the strong-field constraints from black-hole shadows to the weak-field tests, we expand the tt element in powers of $r^{-1}$ as
\begin{equation}
-g_{tt}=1-\frac{2}{r}+2\left(\frac{\bar{\beta}-\bar{\gamma}}{r^2}\right) - 2 \left(\frac{\zeta}{r^3}\right)+{\cal O}\left(r^{-4}\right)\;.
\label{eq:tt_PN2}
\end{equation}
Hereafter, we set $G=c=M=1$, where $G$ is the gravitational constant, $c$ is the speed of light, and $M$ is the black-hole mass. In this equation, we have employed the usual PPN parameters $\bar{\beta}$ and $\bar{\gamma}$ and added a 2PN term parametrized by the quantity $\zeta$. Weak-field tests have placed strong constraints on the 1PN parameters to be equal to unity to within a few parts in $10^5$~\cite{Will2014}. Even though modified gravity theories may not satisfy Birkhoff's theorem and, therefore, the values of the 1PN parameters may be different outside the Sun and outside a black hole, we make here the very conservative assumption that the solar-system limits are applicable to the external spacetimes of astrophysical black holes and set $\bar{\beta}-\bar{\gamma}\simeq 0$. If the $tt$ element of the black-hole metric has indeed a vanishing 1PN term, as required by the weak-field tests, and terminates at the 2PN term, the radius of the circular photon orbit would be
\begin{equation}
r_{\rm ph}=3+\frac{5}{9}\zeta
\end{equation}
and the size of the black-hole shadow would be
\begin{equation}
r_{\rm sh}=3\sqrt{3}\left(1+\frac{1}{9}\zeta\right)\;.
\label{eq:rshPPN}
\end{equation}
This is a quantitative demonstration of the fact that the size of the black-hole shadow probes the behavior of the spacetime  at least at the 2PN order. Moreover, the size of the black-hole shadow depends linearly on the magnitude of the 2PN term.

To explore in detail the constraints imposed by the EHT results, we will consider, as concrete examples of regular, parametric extensions to the Kerr metric, the metrics developed in Refs.~\cite{Johannsen2011,Johannsen2013} (hereafter the JP metric) and in Refs.~\cite{Vigeland2011, Gair2011} (hereafter the MGBK metric). Table~\ref{tab:table1} shows the 1PN and 2PN parameters (see Eq.~\ref{eq:metric_gen}) for these metrics, when the spin parameter is set to zero and only leading orders of the parameters are considered. From the analytic argument above, we expect the shadow sizes to be determined primarily by the parameters that control the 2PN and higher-order terms for these metrics. Hereafter, we define the spin of a given metric as the dimensionless ratio $J/M^2$ of the lowest-order current moment, i.e, the angular momentum, to the square of the lowest-order mass moment, i.e., the Keplerian mass, of the spacetime.

The JP metric has four lowest-order parameters to describe possible deviations from Kerr~\cite{Johannsen2013}. The outlines of black-hole shadows for this metric have been calculated in Refs.~\cite{Johannsen2013c,Medeiros2020} and were shown to depend very weakly on the black-hole spin. Setting the spin to zero, we use Eq.~(\ref{eq:rsh}) for the full metric to derive the shadow size as a function of the deviation parameters. We find that, in this limit, the shadow size depends entirely on one of the deviation parameters, $\alpha_{13}$, which is also the one that controls the 2PN terms of the metric. The complete expression is very complicated to display here but a power-law expansion is
\begin{equation}
r_{\rm sh,JP}=3\sqrt{3}\left[1+\frac{1}{27}\alpha_{13}-\frac{1}{486}\alpha_{13}^2+{\cal O}(\alpha_{13}^3)\right]\;.
\label{eq:rshJP}
\end{equation}
Note that the coefficient of the deviation parameter $\alpha_{13}$ is different from what we would have expected from eq.~(\ref{eq:rshPPN}) because the JP metric does not terminate at the 2PN order. Requiring that the shadow size is consistent to within 17\% with the 2017 EHT measurement for M87 places a bound on the deviation parameter $-3.6<\alpha_{13}<5.9$. The left panel of Fig.~\ref{fig:metrics} shows the corresponding limits on $\alpha_{13}$ obtained numerically from the full JP metric, when the black-hole spin is taken into account and the second metric parameter that affects the shadow size for a spinning black hole, i.e., $\alpha_{22}$,  is varied. As evident here, the constraints on $\alpha_{13}$ change only mildly when effects that introduce deviations from spherical symmetry are included. Therefore, for the JP metric, the EHT measurement  constrains predominantly the deviation parameter $\alpha_{13}$, which controls the 2PN terms. 

The MGBK metric has four lowest-order parameters to describe possible deviations from Kerr~\cite{Gair2011} without requiring the 1PN deviation to vanish (see Table~\ref{tab:table1}). The outlines of black-hole shadows have been calculated  in Ref.~\cite{Medeiros2020} and their overall sizes were shown to depend primarily on the parameters $\gamma_{3,3}$, $\gamma_{1,2}$, and $\gamma_{4,2}$ (see Fig.~8 of~\cite{Medeiros2020}). In its original formulation, the parameter $\gamma_{3,3}$ describes frame dragging in a manner that remains finite even for nonspinning black holes (see Eq.~[17] of \cite{Gair2011}). Here, we scale this parameter with spin, i.e., write $\gamma^\prime_{3,3}=\gamma_{3,3} a$ to remove the divergent behavior of the shadow size with $a\rightarrow 0$ found in Ref.~\cite{Medeiros2020}. We also set  $\gamma_{4,2}=-\gamma_{1,2}/2$ for this metric to be consistent with Solar System tests at the 1PN order. In this case, the magnitude of potential 2PN deviations becomes equal to $\zeta_{\rm MGBK}=\gamma_{1,2}$.

With these redefinitions, the size of the shadow for the MGBK metric depends primarily on parameter $\gamma_{1,2}$ and only weakly on spin. As before, we calculate analytically the shadow size for this metric using Eq.~(\ref{eq:rsh}) having set the spin equal to zero. We again display only an expansion of the size in the deviation parameter $\gamma_{1,2}$:
\begin{equation}
r_{\rm sh,MGBK}=3\sqrt{3}\left[1+\frac{1}{27}\gamma_{1,2}+{\cal O}(\gamma_{1,2}^3)\right]\;.
\label{eq:rshMGBK}
\end{equation}
Requiring that the shadow size is consistent to within 17\% with the 2017 EHT measurement for M87 places a bound on the deviation parameter $-5.0<\gamma_{1,2}<4.9$.  The right panel of Fig.~\ref{fig:metrics} shows the corresponding constraints obtained numerically from the full solution, when the black-hole spin is taken into account and the other deviation parameters are varied. Again, the constraints on $\gamma_{1,2}$ change only mildly when effects that introduce deviations from spherical symmetry are included.

Even though the complex functional forms of the various elements in the two metrics we considered here are 
very different from each other, in both cases the predicted size of the black-hole shadow depends almost 
exclusively (and in a very similar manner) on the deviation parameter that controls the 2PN and higher-order terms for each metric.
This conclusion remains the same when we use, e.g., the RZ metric [29], for which the deviations from Kerr are introduced by a sequence of parameters, with $a_i$ controlling primarily the $i+1$ PN order. For this metric, $\zeta=-4 \alpha_1$ and requiring that the predicted shadow size is consistent with the EHT measurements leads to the constraint $-1.2<\alpha_1<1.3$. This supports our conclusion that an EHT measurement of the size of a black hole leads to metric tests that are inaccessible to weak-field tests. 

In this Letter we have allowed for only one of the high-order PN parameters of the $g_{tt}$ component of each metric to deviate from its Kerr value in order to show that significant constraints can be obtained even with the current EHT results. However, if more than one PN parameters of the same metric component are included, then the size measurement of the black-hole shadow will instead lead to a constraint on a linear combination of these parameters. Similar constraints will be possible in the very near future with EHT observations of the black hole in the center of the Milky Way, for which there is no ambiguity in the inferred mass. In that case, monitoring of individual stellar orbits has provided very precise measurements of its mass-to-distance ratio~\cite{Gravity2019,*Gravity2020,*Do2019} leading to a prediction of $47-53~\mu$as for its shadow diameter, depending on the black-hole spin.

Observations of double neutron stars~\cite{Wex2014} and of coalescing black holes with LIGO/VIRGO~\cite{Abbott2016} also probe the strong-field properties of their gravitational fields and lead to post-Newtonian constraints of similar magnitude as the ones we obtain here. The mass and curvature scale of the stellar-mass sources are eight orders of magnitude different from those of the M87 black hole, thereby probing a very different regime of gravitational parameters~\cite{Psaltis2019, PaperI}. It is this combination of gravitational tests across different scales that will provide complementary and comprehensive constraints on possible modifications of the fundamental gravitational theory.

\bigskip

\begin{acknowledgments}
The authors of the present paper thank the following
organizations and programs: the Academy
of Finland (projects 274477, 284495, 312496); the
Advanced European Network of E-infrastructures for
Astronomy with the SKA (AENEAS) project, supported
by the European Commission Framework Programme
Horizon 2020 Research and Innovation action
under grant agreement 731016; the Alexander
von Humboldt Stiftung; the Black Hole Initiative at
Harvard University, through a grant (60477) from
the John Templeton Foundation; the China Scholarship
Council; Comisión Nacional de Investigación
Científica y Tecnológica (CONICYT, Chile, via PIA
ACT172033, Fondecyt projects 1171506 and 3190878,
BASAL AFB-170002, ALMA-conicyt 31140007); Consejo
Nacional de Ciencia y Tecnología (CONACYT,
Mexico, projects 104497, 275201, 279006, 281692);
the Delaney Family via the Delaney Family John A.
Wheeler Chair at Perimeter Institute; Dirección General
de Asuntos del Personal Académico-—Universidad
Nacional Autónoma de México (DGAPA-—UNAM,
project IN112417); the European Research Council Synergy
Grant "BlackHoleCam: Imaging the Event Horizon
of Black Holes" (grant 610058); the Generalitat
Valenciana postdoctoral grant APOSTD/2018/177 and
GenT Program (project CIDEGENT/2018/021); the
Gordon and Betty Moore Foundation (grants GBMF-
3561, GBMF-5278); the Istituto Nazionale di Fisica
Nucleare (INFN) sezione di Napoli, iniziative specifiche
TEONGRAV; the International Max Planck Research
School for Astronomy and Astrophysics at the
Universities of Bonn and Cologne; the Jansky Fellowship
program of the National Radio Astronomy Observatory
(NRAO); the Japanese Government (Monbukagakusho:
MEXT) Scholarship; the Japan Society for
the Promotion of Science (JSPS) Grant-in-Aid for JSPS
Research Fellowship (JP17J08829); the Key Research
Program of Frontier Sciences, Chinese Academy of
Sciences (CAS, grants QYZDJ-SSW-SLH057, QYZDJSSW-
SYS008, ZDBS-LY-SLH011); the Leverhulme Trust Early Career Research
Fellowship; the Max-Planck-Gesellschaft (MPG);
the Max Planck Partner Group of the MPG and the
CAS; the MEXT/JSPS KAKENHI (grants 18KK0090,
JP18K13594, JP18K03656, JP18H03721, 18K03709,
18H01245, 25120007); the MIT International Science
and Technology Initiatives (MISTI) Funds; the Ministry
of Science and Technology (MOST) of Taiwan (105-
2112-M-001-025-MY3, 106-2112-M-001-011, 106-2119-
M-001-027, 107-2119-M-001-017, 107-2119-M-001-020,
and 107-2119-M-110-005); the National Aeronautics and
Space Administration (NASA, Fermi Guest Investigator
grant 80NSSC17K0649 and Hubble Fellowship grant
HST-HF2-51431.001-A awarded by the Space Telescope
Science Institute, which is operated by the Association
of Universities for Research in Astronomy, Inc.,
for NASA, under contract NAS5-26555); the National
Institute of Natural Sciences (NINS) of Japan; the National
Key Research and Development Program of China
(grant 2016YFA0400704, 2016YFA0400702); the National
Science Foundation (NSF, grants AST-0096454,
AST-0352953, AST-0521233, AST-0705062, AST-
0905844, AST-0922984, AST-1126433, AST-1140030,
DGE-1144085, AST-1207704, AST-1207730, AST-
1207752, MRI-1228509, OPP-1248097, AST-1310896,
AST-1312651, AST-1337663, AST-1440254, AST-
1555365, AST-1715061, AST-1615796, AST-1716327,
OISE-1743747, AST-1816420); the Natural Science
Foundation of China (grants 11573051, 11633006,
11650110427, 10625314, 11721303, 11725312, 11933007); the Natural
Sciences and Engineering Research Council of
Canada (NSERC, including a Discovery Grant and
the NSERC Alexander Graham Bell Canada Graduate
Scholarships-Doctoral Program); the National Youth
Thousand Talents Program of China; the National Research
Foundation of Korea (the Global PhD Fellowship
Grant: grants NRF-2015H1A2A1033752, 2015-
R1D1A1A01056807, the Korea Research Fellowship Program:
NRF-2015H1D3A1066561); the Netherlands Organization
for Scientific Research (NWO) VICI award
(grant 639.043.513) and Spinoza Prize SPI 78-409; the
New Scientific Frontiers with Precision Radio Interferometry
Fellowship awarded by the South African Radio
Astronomy Observatory (SARAO), which is a facility
of the National Research Foundation (NRF), an
agency of the Department of Science and Technology
(DST) of South Africa; the Onsala Space Observatory
(OSO) national infrastructure, for the provisioning
of its facilities/observational support (OSO receives
funding through the Swedish Research Council under
grant 2017-00648) the Perimeter Institute for Theoretical
Physics (research at Perimeter Institute is supported
by the Government of Canada through the Department
of Innovation, Science and Economic Development
and by the Province of Ontario through the
Ministry of Research, Innovation and Science); the Russian
Science Foundation (grant 17-12-01029); the Spanish
Ministerio de Economía y Competitividad (grants
AYA2015-63939-C2-1-P, AYA2016-80889-P, PID2019-108995GB-C21); the State
Agency for Research of the Spanish MCIU through
the "Center of Excellence Severo Ochoa" award for
the Instituto de Astrofísica de Andalucía (SEV-2017-
0709); the Toray Science Foundation; the Consejería de Economía, Conocimiento, Empresas y Universidad of the Junta de Andalucía (grant P18-FR-1769), the Consejo Superior de Investigaciones Científicas (grant 2019AEP112);
the US Department
of Energy (USDOE) through the Los Alamos National
Laboratory (operated by Triad National Security,
LLC, for the National Nuclear Security Administration
of the USDOE (Contract 89233218CNA000001));
the Italian Ministero dell’Istruzione Università e Ricerca
through the grant Progetti Premiali 2012-iALMA (CUP
C52I13000140001); the European Union’s Horizon 2020
research and innovation programme under grant agreement
No 730562 RadioNet; ALMA North America Development
Fund; the Academia Sinica; Chandra TM6-
17006X; the GenT Program (Generalitat Valenciana)
Project CIDEGENT/2018/021. This work used the
Extreme Science and Engineering Discovery Environment
(XSEDE), supported by NSF grant ACI-1548562,
and CyVerse, supported by NSF grants DBI-0735191,
DBI-1265383, and DBI-1743442. We thank
the staff at the participating observatories, correlation
centers, and institutions for their enthusiastic support.
ALMA is a partnership of the European Southern Observatory (ESO;
Europe, representing its member states), NSF, and
National Institutes of Natural Sciences of Japan, together
with National Research Council (Canada), Ministry
of Science and Technology (MOST; Taiwan),
Academia Sinica Institute of Astronomy and Astrophysics
(ASIAA; Taiwan), and Korea Astronomy and
Space Science Institute (KASI; Republic of Korea), in
cooperation with the Republic of Chile. The Joint
ALMA Observatory is operated by ESO, Associated
Universities, Inc. (AUI)/NRAO, and the National Astronomical
Observatory of Japan (NAOJ). The NRAO
is a facility of the NSF operated under cooperative agreement
by AUI. APEX is a collaboration between the
Max-Planck-Institut fur Radioastronomie (Germany),
ESO, and the Onsala Space Observatory (Sweden). The
SMA is a joint project between the SAO and ASIAA
and is funded by the Smithsonian Institution and the
Academia Sinica. The JCMT is operated by the East
Asian Observatory on behalf of the NAOJ, ASIAA, and
KASI, as well as the Ministry of Finance of China, Chinese
Academy of Sciences, and the National Key R\&D
Program (No. 2017YFA0402700) of China. Additional
funding support for the JCMT is provided by the Science
and Technologies Facility Council (UK) and participating
universities in the UK and Canada. The
LMT is a project operated by the Instituto Nacional
de Astrofisica, Optica, y Electronica (Mexico) and the
University of Massachusetts at Amherst (USA). The
IRAM 30-m telescope on Pico Veleta, Spain is operated
by IRAM and supported by CNRS (Centre National de
la Recherche Scientifique, France), MPG (Max-Planck-
Gesellschaft, Germany) and IGN (Instituto Geográfico
Nacional, Spain). The SMT is operated by the Arizona
Radio Observatory, a part of the Steward Observatory
of the University of Arizona, with financial support of
operations from the State of Arizona and financial support
for instrumentation development from the NSF.
The SPT is supported by the National Science Foundation
through grant PLR- 1248097. Partial support is
also provided by the NSF Physics Frontier Center grant
PHY-1125897 to the Kavli Institute of Cosmological
Physics at the University of Chicago, the Kavli Foundation
and the Gordon and Betty Moore Foundation grant
GBMF 947. The SPT hydrogen maser was provided on
loan from the GLT, courtesy of ASIAA. The EHTC has
received generous donations of FPGA chips from Xilinx
Inc., under the Xilinx University Program. The EHTC
has benefited from technology shared under open-source
license by the Collaboration for Astronomy Signal Processing
and Electronics Research (CASPER). The EHT
project is grateful to T4Science and Microsemi for their
assistance with Hydrogen Masers. This research has
made use of NASA’s Astrophysics Data System. We
gratefully acknowledge the support provided by the extended
staff of the ALMA, both from the inception of
the ALMA Phasing Project through the observational
campaigns of 2017 and 2018. We would like to thank
A. Deller and W. Brisken for EHT-specific support with
the use of DiFX. We acknowledge the significance that
Maunakea, where the SMA and JCMT EHT stations
are located, has for the indigenous Hawaiian people.\end{acknowledgments}

\bibliography{prl.bib}
\end{document}